\documentclass[aps,preprint,pre,eqsecnum,longbibliography]{revtex4-1} 
\usepackage{natbib}
\usepackage{soul}
\usepackage{amsfonts}
\usepackage{amsmath}
\usepackage{amssymb}
\usepackage{graphicx}
\usepackage{mathrsfs}  
\usepackage{color} 
\usepackage{gensymb} 
\providecommand\bnabla{\boldsymbol{\nabla}}

\providecommand\bcdot{\boldsymbol{\cdot}}

\providecommand\bi{\mathbf{i}}

\providecommand\bu{\mathbf{u}}

\providecommand\bx{\mathbf{x}}
\providecommand\by{\mathbf{y}}

\providecommand\bzero{\mathbf{0}}

\providecommand\be{\mathbf{\hat{e}}}
\providecommand\bn{\mathbf{\hat{n}}}

\newcommand{\pd}[2]{\frac{\partial #1}{\partial #2}}

\newcommand{\ub}[1]{^{({#1})}}

\begin{document} 
\begin{abstract}
We use matched asymptotics to derive analytical formulae for the acoustic impedance of a subwavelength orifice consisting of a cylindrical perforation in a rigid plate. In the inviscid case, an end correction to the length of the orifice due to Rayleigh is shown to constitute an exponentially accurate approximation in the limit where the aspect ratio of the orifice is large; in the opposite limit, we derive an algebraically accurate correction, depending upon the logarithm of the aspect ratio, to the impedance of a circular aperture in a zero-thickness screen. Viscous effects are considered in the limit of thin Stokes boundary layers, where a boundary layer analysis in conjunction with a reciprocity argument provides the perturbation to the impedance as a quadrature of the basic inviscid flow. We show that for large aspect ratios the latter perturbation can be captured with exponential accuracy by introducing a second end correction whose value is calculated to be in between two guesses commonly used in the literature; we also derive an algebraically accurate approximation in the small-aspect-ratio limit. The viscous theory reveals  that the resistance exhibits a minimum as a function of aspect ratio, with the orifice radius held fixed. It is evident that the resistance grows in the long-aspect-ratio limit; in the opposite limit, resistance is amplified owing to the large velocities close to the sharp edge of the orifice.
The latter amplification arrests only when the the plate is as thin as the Stokes boundary layer. The analytical approximations derived in this paper could be used to improve circuit modelling of resonating acoustic devices.  
\end{abstract}

\title[Impedance of a cylindrical orifice]
{Acoustic impedance of a cylindrical orifice}
\author{Rodolfo Brand{\~a}o} \author{Ory Schnitzer}

\affiliation{Department of Mathematics, Imperial College London, London SW7 2AZ, UK}

\date{\today}

\title[Impedance of a cylindrical orifice]
{Acoustic impedance of a cylindrical orifice}
\maketitle

\section{Introduction and summary of main results}
\subsection{Background}
The acoustic impedance of a perforation is defined as 
\begin{equation}\label{Z}
\mathcal{Z}={\mathcal{P}}/{\mathcal{Q}},
\end{equation}
where $\mathcal{P}$ and $\mathcal{Q}$ are respectively the phasors representing the pressure drop and volumetric flux across and through the perforation \citep{morse:1986}. The pressure drop $\mathcal{P}$ is well defined in the limit where the perforation is small compared to the acoustic wavelength. In this setting, the flow on the scale of the perforation is incompressible and the acoustic pressure is spatially uniform on scales intermediate between the perforation and the wavelength \citep{crighton:1992}. The importance of determining the acoustic impedance of a perforation as a function of frequency and geometry stems from the long-standing 
 popularity of circuit models in the design of structured acoustic devices \citep{zwikker:1949,ingard:1953,melling:1973,maa:1998} and metamaterials \citep{fang:2006,zhang:2009,li:2016,jimenez:2016}. 

\begin{figure}
\begin{center}
\includegraphics[scale=0.6]{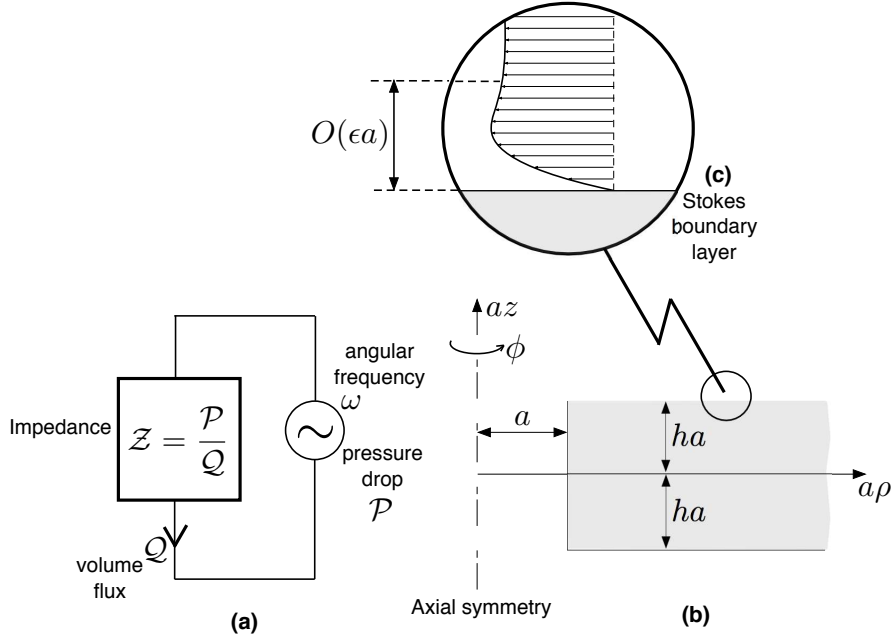}
\caption{Impedance model (a) of a subwavelength cylindrical orifice (b). Viscous effects are considered in the limit of thin Stokes boundary layers (c).}
\label{fig:sketch}
\end{center}
\end{figure}

In this paper, we systematically derive accurate analytical approximations for the acoustic impedance of an orifice consisting of a cylindrical perforation of radius $a$ in a rigid infinite plate of thickness $2ha$, focusing on the linear regime of small-amplitude oscillations without bias flow (see figure \ref{fig:sketch}). This is the setting considered in some of the most classical works on the subject, including analyses by Lord Rayleigh \cite{rayleigh:1871,rayleigh:1896} of the inviscid problem and later extensions by Sivian \cite{sivian:1935} and Ingard \cite{ingard:1953} to include viscous effects. 
Our analysis is based on the method of matched asymptotic expansions \citep{hinch:1991,crighton:1992} in the limits of large and small orifice aspect ratio $h$, with viscous effects studied in the limit of thin Stokes boundary layers. We preface our analysis with a discussion of our main results in the context of existing approximations. 

\subsection{Inviscid case}
When viscous effects can be neglected, the flow through the perforation is irrotational, oscillating with a $\pi/2$ phase difference relative to the pressure field. In that case, the impedance is strictly imaginary and it is traditional to instead consider the Rayleigh conductivity \citep{howe:1998}:
\begin{equation}\label{K def}
\mathcal{K}=-{i\varrho\omega\mathcal{Q}}/{\mathcal{P}},  
\end{equation}
where we denote by $\varrho$  the density of the fluid and $\omega$ the angular frequency. Assuming the convention where flux is measured in the direction of pressure drop and phasor quantities rotate clockwise, $\mathcal{K}$ is real and positive. Note that $\mathcal{K}$ does not actually correspond to an acoustic conductivity; rather, it is proportional to the susceptance --- the inverse of the reactance $\operatorname{Im}\mathcal{Z}$. The term ``conductivity'' was used in this context by Rayleigh in light of a mathematical analogy with an electrical problem (see appendix \ref{app:anal}). 

While an exact expression for $\mathcal{K}$ is unknown for a general cylindrical orifice, Lord Rayleigh \cite{rayleigh:1871,rayleigh:1896} provides useful limits, approximations and bounds. In particular, for $h=0$ the potential flow can be solved in closed form, giving $\mathcal{K}=2a$. In the opposite limit, end effects are negligible and assuming a developed solution in the channel part of the orifice readily gives $\mathcal{K}\approx \pi a/2h$. For arbitrary $h$, Rayleigh gives the lower and upper bounds
\begin{equation}\label{bounds}
\frac{\pi a}{2h+\frac{16}{3\pi}}<{\mathcal{K}}<\frac{\pi a}{2h+\frac{\pi}{2}},
\end{equation}
which agree with the limiting cases mentioned above. Alluding to the large-aspect-ratio limit $h\gg1$, it is customary to recast $\mathcal{K}$ in terms of an effective end correction $2a\Delta h$ to the channel length:
\begin{equation}\label{delta} 
\mathcal{K}=\frac{\pi a}{2(h+\Delta h)},  
\end{equation}
with \eqref{bounds} implying ${\pi}/{4}<\Delta h < {8}/{3\pi}$. As noted by Rayleigh, it is only strictly correct to  interpret \eqref{delta} as an exact relation if $\Delta h$ is allowed to vary with $h$; in practice, however, the use of \eqref{delta} in conjunction with a constant $\Delta h$ is ubiquitous. A popular choice is to use the upper bound $8/3\pi\approx0.8488$ \citep{ingard:1953,kinsler:1999}. Alternatively, Rayleigh gives a recipe to calculate $\Delta h$ for large $h$; his recipe is based on an intuitive decomposition of the fluid domain into a channel region, where the flow is developed, and aperture regions governed by a canonical potential problem. Solving that canonical problem gives $\Delta h\approx 0.8217$ \citep{rayleigh:1871,daniell1:1915,daniell2:1915,king:1936,nomura:1960,norris:1989,wendoloski:93}. 
Surprisingly, \eqref{delta} with Rayleigh's $\Delta h$ predicts $\mathcal{K}$ accurately down to around $h\approx 0.1$ \citep{laurens:2013}. 
\begin{figure}
\begin{center}
\includegraphics[scale=0.266]{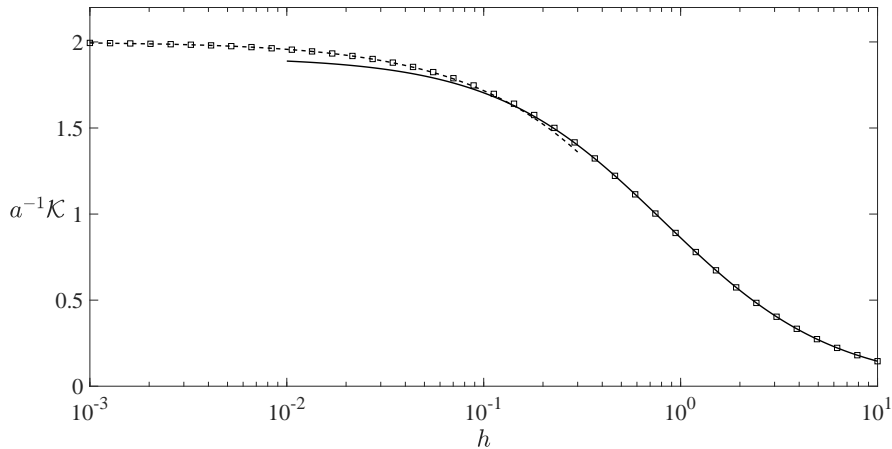}
\caption{The Rayleigh conductivity $\mathcal{K}$ in the inviscid case, normalised by the orifice radius, as a function of the aspect ratio $h$. Solid line: large-$h$ approximation \eqref{long inv}. Dashed line: small-$h$ approximation \eqref{short inv}. Symbols: numerical solution of the inviscid problem. }
\label{fig:K0}
\end{center}
\end{figure}

We next summarise our main results pertaining to the inviscid problem. In the large-aspect-ratio limit, our analysis effectively formalises Rayleigh's decomposition using the method of matched asymptotic expansions, revealing that Rayleigh's recipe for a constant end correction constitutes an asymptotic approximation with an error that is beyond all orders. Namely, we find that 
\begin{equation}\label{long inv}
\mathcal{K}/a\sim\frac{\pi}{2(h+l)}+e.s.t. \quad \text{as} \quad h\to\infty,
\end{equation}
where $l\approx 0.8217$ and $e.s.t.~$stands for exponentially small terms in the indicated limit. In the opposite small-aspect-ratio limit, we match an outer region, where the geometry is approximately that of a circular aperture in a zero-thickness screen, and an inner region close to the edge of the screen. We thereby find
\begin{equation}\label{short inv}
\mathcal{K}/a\sim 2 - \frac{2h}{\pi}\left(\ln\frac{\pi}{h}+1\right) + \cdots \quad \text{as} \quad h\to0,
\end{equation}
with an algebraic relative error, namely smaller than some positive power of $h$. The appearance of a logarithm is linked to the displacement of the singular outer flow around the sharp edge of the orifice owing to its finite thickness \citep{morse:1986}. An analogous logarithm occurs in the two-dimensional problem of plane-wave diffraction from a semi-infinite plate whose thickness is small compared to the wavelength \citep{crighton:1973}. 
In figure \ref{fig:K0} we compare the asymptotic approximations \eqref{long inv} and \eqref{short inv} against values of $a^{-1}\mathcal{K}$ calculated numerically as described in \S\ref{sec:form}. 

\subsection{Viscous effects}\label{ssec:viscous}
Viscous effects on the impedance of a cylindrical orifice are characterised by the dimensionless parameter $\epsilon\! =\!a^{-1}\sqrt{\mu/\varrho\omega}$, $\mu$ being the fluid viscosity, which characterises the thickness of the Stokes boundary layer in comparison with the orifice radius (see figure \ref{fig:sketch}(c)). The inviscid regime previously considered corresponds to the limiting case $\epsilon\!=\!0$; the opposite extreme $\epsilon\!=\!\infty$  corresponds to the creeping flow regime where the impedance is purely resistive \citep{sampson:1891,dagan:1982}. Our interest here is in the thin-boundary-layer limit $\epsilon\!\ll\!1$, which corresponds to the regime most relevant to acoustic resonators \citep{ingard:1953}; indeed, while the viscous correction to the impedance is naturally small in this case, it constitutes a dominant contribution to the resistance. It should be noted that in the subwavelength regime considered in this paper, thermal dissipative effects are typically negligible \citep{ingard:1953,stinson:1985}. 

Most models of acoustic impedance that account for viscous effects are ultimately based on the infinite-channel theories of Lord Rayleigh \cite{rayleigh:1896} and Crandall \cite{crandall:1926}, which for small $\epsilon$ give \citep{melling:1973}
\begin{equation}\label{melling}
\frac{\mathcal{Z}a}{\varrho \omega}\approx 2h \times \left\{-\frac{i}{\pi}+\epsilon\frac{(1-i)\sqrt{2}}{\pi}\right\}.\end{equation}
Analogously to the inviscid case, there have been attempts to extend \eqref{melling} to arbitrary $h$ by introducing end corrections. In particular, Sivian  \cite{sivian:1935} boldly proposed to use the same end correction as in the inviscid case, namely to replace $h$ in \eqref{melling} by $h+\Delta h$. Another popular choice suggested by Ingard \cite{ingard:1953} is to separate \eqref{melling} into inviscid and viscous contributions, which are respectively corrected by incrementing $h$ by $\Delta h$ and $(\Delta h)_v=1$. While such end corrections are clearly heuristic, they have been shown to give reasonably good agreement with experiments, even for moderate values of $h$ \citep{thurston:1953,ingard:1953,stinson:1985,peat:2010,komkin:2017}.

Let us now summarise our main results pertaining to the viscous problem. To leading order in $\epsilon$, the flow through the orifice is identical to that in the inviscid case, excluding an $O(\epsilon)$-thick Stokes boundary layer adjacent to the orifice walls. To first order in $\epsilon$, the flow outside the boundary layer remains inviscid; it is perturbed, however, owing to the displacement of the basic inviscid flow  by the Stokes layer. No need arises to solve for the perturbed inviscid flow; rather, a reciprocity relation between the basic and perturbed flows explicitly provides the $O(\epsilon)$ correction to the impedance. We thereby find
\begin{equation}\label{viscous impedance}
\frac{\mathcal{Z}a}{\varrho\omega}\sim -\frac{ia}{\left.\mathcal{K}\right|_{\epsilon=0}} + \epsilon\Theta\frac{1-i}{\sqrt{2}}+\cdots \quad \text{as} \quad \epsilon\to0,
\end{equation} 
where $\left.\mathcal{K}\right|_{\epsilon=0}$ is the Rayleigh conductivity in the inviscid case and $\Theta$ is a real-valued function of $h$ provided as a quadrature of the basic inviscid flow. 
\begin{figure}
\begin{center}
\includegraphics[scale=0.266]{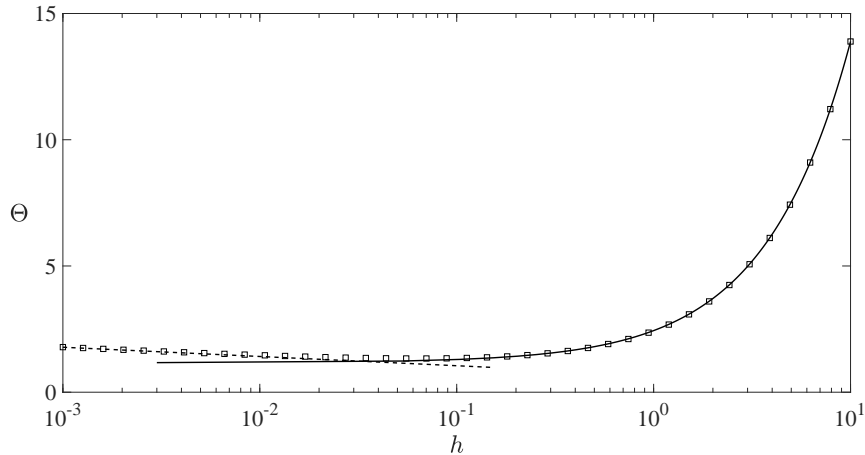}
\caption{The parameter $\Theta$, which defines the viscous correction to the acoustic impedance of the orifice via \eqref{viscous impedance}, as a function of the aspect ratio $h$. Solid line: large-$h$ approximation \eqref{long v}. Dashed line: small-$h$ approximation \eqref{short v}. Symbols: \eqref{Theta def} evaluated using the numerical solution of the inviscid problem.}
\label{fig:theta}
\end{center}
\end{figure}

The function $\Theta$ is evaluated in the large- and small-aspect-ratio limits using the matched-asymptotic solutions of the basic inviscid flow. In the former limit, we find  
\begin{equation}\label{long v}
\Theta\sim \frac{4}{\pi}(h+l_v) + e.s.t. \quad \text{as} \quad h\to\infty,
\end{equation}
where ${l_v}\approx 0.91$ is obtained from the same canonical aperture problem defining the inviscid end correction $l$. We note that $l_v$ can be interpreted as the ``correct'' value of $(\Delta h)_v$; incidentally, it is nearly in the middle between the values guessed by Sivian \cite{sivian:1935} and Ingard \cite{ingard:1953}. In the small-aspect-ratio limit we find 
\begin{equation}\label{short v}
\Theta\sim \frac{1}{2}\left(\frac{1}{\pi}\ln\frac{\pi}{h}+1\right) + \cdots \quad \text{as} \quad h\to0,
\end{equation}
with an algebraic relative error. The logarithm here is linked to enhanced dissipation close to the sharp edge of the orifice; previously, Morse and Ingard \cite{morse:1986} derived the leading logarithmic term in \eqref{short v}. In figure \ref{fig:theta} we compare the asymptotic approximations \eqref{long v} and \eqref{short v} against values of $\Theta$ calculated numerically as described in \S\ref{sec:viscous}. Observe that $\Theta$ exhibits a minimum as a function of $h$; this is also evident from \eqref{long v} and \eqref{short v}.  

In the above theory, the resistance $\operatorname{Re}\mathcal{Z}$ diverges logarithmically as $h\to0$ with $\epsilon$ fixed [cf.~\eqref{viscous impedance} and \eqref{short v}]. Note, however, that this divergence was obtained in the double limit $\epsilon\!\to\!0$ followed by $h\!\to\!0$; physically, this corresponds to the case $h\!\gg\!\epsilon$. The divergence of the resistance arrests when $h$ becomes commensurate with $\epsilon$ and in that case a coarse logarithmic approximation can be deduced as 
\begin{equation}\label{intro special}
\frac{\mathcal{Z}a}{\varrho\omega}\sim -\frac{i}{2} - \epsilon\left(\frac{i-1}{\sqrt{2}} + i\frac{h}{\epsilon}\right)\ln\frac{1}{\epsilon}+\cdots \quad \text{as}  \quad \epsilon\to 0 \quad \text{with} \quad h/\epsilon \quad \text{fixed.}
\end{equation}

The remainder of this paper is dedicated to the derivation of the results stated in this introduction. We begin in \S\ref{sec:form} by formulating the inviscid problem, which is then analysed in the large- and small-$h$ limits in \S\ref{sec:long} and \S\ref{sec:short}, respectively. The viscous analysis is carried out in \S\ref{sec:viscous}. 

\section{Formulation of the inviscid problem}\label{sec:form}
Henceforth we assume a dimensionless convention where lengths are normalised by the orifice radius $a$ and the phasor fields $p$ and $\bu$ give the the physical pressure and velocity fields as 
\refstepcounter{equation}
$$
\mathcal{P}\operatorname{Re}\left\{p e^{-i\omega t}\right\}, \quad \frac{\mathcal{P}}{\varrho\omega a}\operatorname{Re}\left\{\bu e^{-i\omega t}\right\},
\label{velocity and pressure}
\eqno{(\theequation{\mathit{a},\mathit{b}})}
$$
respectively, where $t$ denotes time. We also introduce the dimensionless cylindrical coordinates $(\rho,z,\phi)$,  which are defined as shown in figure \ref{fig:sketch}(b); the unit vectors associated with these coordinates are $(\be_{\rho},\be_z,\be_{\phi})$. 

Our interest is in the case where the dimensions of the orifice are sufficiently small relative to the wavelength so that compressibility effects are negligible \citep{rayleigh:1896}. When dissipative effects are also negligible (see \S\ref{sec:viscous}), the equations governing the oscillatory flow through the orifice reduce to 
\refstepcounter{equation}
$$
\bu=-i\bnabla p, \quad \bnabla\bcdot\bu=0. \label{relations}
\eqno{(\theequation{\mathit{a},\mathit{b}})}
$$
These can be combined to give Laplace's equation for the pressure field:
\begin{equation}\label{lap}
\nabla^2p = 0.
\end{equation}
On the rigid boundary of the orifice, impermeability  and \eqref{relations}(a) imply the Neumann boundary condition
\begin{equation}\label{neu}
\bn\bcdot\bnabla p=0,
\end{equation}
wherein $\bn$ is the normal unit vector pointing into the fluid domain. The problem governing the scalar field $p$ is closed by far-field conditions representing the oscillating pressure drop across the orifice; without loss of generality, these can be prescribed symmetrically: 
\begin{equation}\label{far}
p\to\pm \frac{1}{2} \quad \text{as} \quad r\to\infty \quad (z\gtrless0),
\end{equation}
where $r\!=\!\sqrt{\rho^2+z^2}$; since $\mathcal{P}$ in \eqref{velocity and pressure} is real, \eqref{far} implies that phase is measured relative to the peak pressure drop in the $-z$ direction. It is evident from the above problem that $p$ is independent of $\phi$ and odd in $z$. 

Our goal is to calculate the Rayleigh conductivity of the orifice [cf.~\eqref{K def}]. Dimensional analysis shows that it possesses the form 
\begin{equation}\label{K def 2}
\mathcal{K}=a{K}(h),
\end{equation}
where $K$ is a dimensionless function of the aspect ratio $h$. The latter function can be calculated as 
\begin{equation}\label{K integral}
K =-i\iint \bn \bcdot \bu\,  dA,
\end{equation}
where the surface integral is over, say, the orifice cross section at $z\!=\!0$ with $\bn\!=\!-\be_z$; it is readily shown from the above formulation that $K$ is real and positive. Since the fluid is incompressible and the orifice boundary is impermeable [cf.~\eqref{relations} and \eqref{neu}], the integration surface can be arbitrarily deformed as long as its boundary remains in contact with the orifice walls. In particular, consider the integration surface to be the intersection of the hemisphere $r\!=\!r^{*}$ (in $z\!>\!0$) and the fluid domain, with $\bn$ pointing towards the origin. By considering the limit $r^{*}\!\!\to\!\infty$, $K$ can be related to the monopole strength in the far-field expansion of $p$ [cf.~\eqref{far}]:
\begin{equation}\label{K in far} 
p \sim \pm \frac{1}{2}\mp \frac{K}{2\pi r} + o(1/r) \quad \text{as} \quad r\to\infty \quad  (z\gtrless0).
\end{equation}

The next two sections are devoted to an asymptotic analysis of the inviscid problem \eqref{lap}--\eqref{far} in the diametric limits of large and small $h$, leading to accurate analytical approximations for $K$. For the sake of validating our analytical  expressions, we have also solved the inviscid problem numerically using a standard mode-matching scheme \citep{king:1936,wendoloski:93}, as well as using Matlab's finite-element Partial Differential Equations Toolbox \cite{MATLAB:2019}. 

\section{Large aspect ratio}\label{sec:long}
\subsection{Leading-order approximation}
It is illuminating to preface our asymptotic analysis in the large-aspect-ratio limit $h\!\to\!\infty$ with an intuitive leading-order derivation. Within the orifice channel, assumed long relative to the aperture radius, the flow is approximately developed. As a consequence, the pressure is cross-sectionally uniform and varies linearly with $z$; thus, neglecting the pressure variations near the apertures, namely setting $p\!=\!\pm1/2$ at $z=\!\pm\!h$, gives $p\!\approx \!z/2h$ and $\bu\!\approx\! -i\be_z/2h$. Substituting this velocity field into \eqref{K integral} yields
\begin{equation}\label{lo long}
K\sim \frac{\pi}{2h} + \ldots \quad \text{as} \quad h\to\infty.
\end{equation}

\begin{figure}
\begin{center}
\includegraphics[scale=0.55]{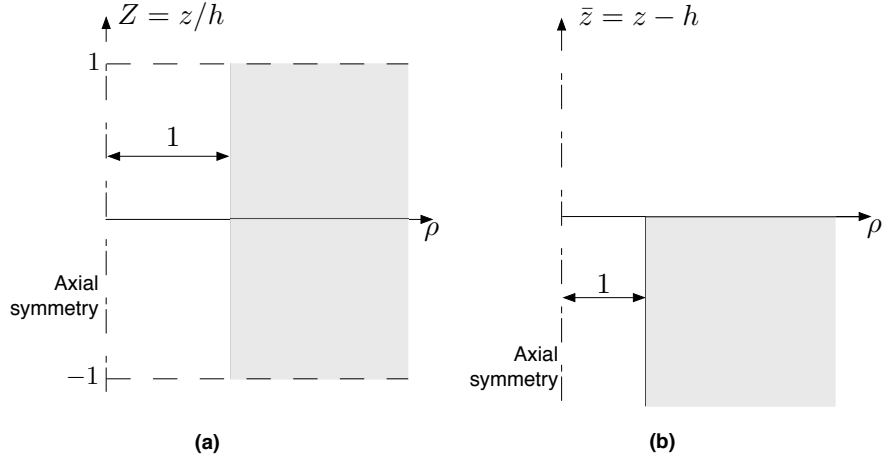}
\caption{Asymptotic regions considered in the large-aspect-ratio limit $h\!\to\!\infty$. (a) Channel region. (b) Aperture region.}
\label{fig:long}
\end{center}
\end{figure}

In what follows, we systematically confirm \eqref{lo long} and extend this expansion to all algebraic orders. To this end, we shall employ the method of matched asymptotic expansions, conceptually decomposing the fluid domain into three asymptotic regions: a channel region and two aperture regions (see figure \ref{fig:long}).

\subsection{Channel region}
The channel region of the orifice is studied by considering the limit $h\!\to\!\infty$ with the stretched coordinates $\rho$ and $Z\!=\!z/h\in(-1,1)$ fixed. As shown in figure \ref{fig:long}(a), in terms of these coordinates the associated fluid domain consists of a finite cylinder of height $2$ and unity radius. Defining  the pressure in this region as $p(\rho,z)\!=\!P(\rho,Z)$, we find from \eqref{lap}--\eqref{far} that $P$ satisfies Laplace's equation in the form
\begin{equation}\label{Laplace P}
\frac{1}{\rho}\pd{}{\rho}\left(\rho\pd{P}{\rho}\right)+\frac{1}{h^2}\pd{^2P}{Z^2}=0,
\end{equation}
the Neumann boundary condition
\begin{equation}\label{Neu P}
\pd{P}{\rho}=0 \quad \text{at} \quad \rho = 1
\end{equation}
and conditions at $Z\!=\!\pm1$ to be determined by asymptotic matching with the aperture regions. 

The general solution to \eqref{Laplace P} and \eqref{Neu P} can be readily written down using separation of variables. Recalling that $P$ is odd in $Z$ and alluding to definition \eqref{K integral}, that solution asymptotically reduces to 
\begin{equation}
\label{Gchannel}
P \sim \frac{h K}{\pi}Z + e.s.t. \quad \text{as} \quad h\to\infty.
\end{equation}
Indeed, non-singular $\rho$-dependent terms vary exponentially with $\pm hZ$; in light of the need to match \eqref{Gchannel} with the aperture regions, where the pressure is of order unity, the latter solutions are necessarily multiplied by exponentially small prefactors.

\subsection{Aperture regions and matching}\label{ssec:aperture}
Using symmetry, we only need to consider a single aperture region. The upper aperture region, say, corresponds to the limit $h\!\to\!\infty$ with the shifted coordinates $\rho$ and $\bar{z}\!=\!z-h$ fixed. The geometry of that region is depicted in figure \ref{fig:long}(b); in particular, note that the orifice cylinder in this limit becomes semi-infinite, with $\bar{z}$ extending to $-\infty$. 

In light of the far-field condition \eqref{far}, we write the pressure in the aperture region as
\begin{equation}\label{pressure shift}
p(\rho,z)=1/2+\bar{p}(\rho,\bar{z}).
\end{equation}
From \eqref{lap}--\eqref{far}, $\bar{p}$ satisfies Laplace's equation
\begin{equation}\label{long lap}
\frac{1}{\rho}\pd{}{\rho}\left(\rho\pd{\bar{p}}{\rho}\right)+\pd{^2\bar{p}}{{\bar{z}}^2}=0,
\end{equation}
the Neumann conditions 
\refstepcounter{equation}
$$
\pd{\bar{p}}{\bar{z}}=0 \quad \text{at} \quad \bar{z}=0 \quad (\rho>1) \quad \text{and} \quad 
\pd{\bar{p}}{\rho}=0 \quad \text{at} \quad \rho=1 \quad  (\bar{z}<0),
\label{long neu}
\eqno{(\theequation{\mathit{a},\mathit{b}})}
$$
the far-field decay condition 
\begin{equation}\label{long far}
\bar{p}\to 0  \quad \text{as} \quad \rho^2+\bar{z}^2\to\infty \quad (\bar{z}>0)
\end{equation}
and matching conditions as $\bar{z}\!\to\!-\infty$. 

The general form of the asymptotic expansion of $\bar{p}$ as $\bar{z}\!\to\!-\infty$ is readily obtained using separation of variables, noting that the geometry is cylindrical in that limit. In principle, that expansion includes exponentially growing terms; such terms, however, must be exponentially small as $h\!\to\!\infty$ for matching with the channel region to be possible. In the absence of exponentially growing terms, matching the net flux through the aperture with that implied by \eqref{Gchannel} enables us to write
\begin{equation}\label{long channel exp}
\bar{p}(\rho,\bar{z};h)\sim \frac{K}{\pi}g(\rho,\bar{z}) + e.s.t. \quad \text{as} \quad h\to\infty,
\end{equation}
where the field ${g}$ satisfies \eqref{long lap}--\eqref{long far} and the matching condition
\begin{equation}\label{g channel pre}
\pd{g}{\bar{z}}\to 1 \quad \text{as} \quad \bar{z}\to-\infty.
\end{equation}

The problem governing $g$ is independent of $h$; it is identical to the canonical aperture problem discussed in the introduction. For matching purposes we note
\begin{equation}\label{g channel}
g\sim \bar{z}-l + e.s.t. \quad \text{as} \quad \bar{z}\to-\infty,
\end{equation}
where $l\!\approx\! 0.8217$ is extracted from a numerical solution of the canonical problem based on the same mode-matching technique mentioned in \S\ref{sec:form}. While the canonical aperture problem has been solved many times in the literature, the goal was always to extract the constant $l$; in \S\ref{sec:viscous} we shall require a second integral property of the solution which has not been calculated previously.

The pressure fields in the channel and aperture regions can now be matched to all algebraic orders using \eqref{Gchannel}, \eqref{pressure shift}, \eqref{long channel exp} and \eqref{g channel}. This gives the following exponentially accurate approximation for the Rayleigh conductivity:
\begin{equation}\label{K exp acc}
K\sim\frac{\pi}{2(h+l)}+e.s.t. \quad \text{as} \quad h\to\infty.
\end{equation}

\section{Small aspect ratio}\label{sec:short}
\subsection{Aperture in a zero-thickness screen}\label{ssec:zerothickness}
In this section we analyse the inviscid problem in the small-aspect-ratio limit $h\!\to\!0$. A leading approximation can be found by simply setting $h\!=\!0$, in which case the orifice degenerates to a circular aperture in a zero-thickness screen. In particular, let $p\!=\!p_0$ and $K\!=\!K_0$ at $h\!=\!0$. From \eqref{lap}--\eqref{far}, $p_0$ satisfies Laplace's equation
\begin{equation}\label{lap thin}
\nabla^2 p_0 = 0,
\end{equation}
the Neumann boundary conditions
\begin{equation}\label{lap neu}
\pd{p_0}{z} = 0 \quad \text{at} \quad z=0 \quad (\rho>1)
\end{equation}
and the far-field conditions
\begin{equation}\label{far thin}
p_0\to \pm\frac{1}{2} \quad \text{as} \quad r\to\infty \quad (z\gtrless0).
\end{equation}
The corresponding value of $K_0$ follows from \eqref{K integral} with $\bu\!=\!-i\bnabla p_0$. 
\begin{figure}
\begin{center}
\includegraphics[scale=0.55]{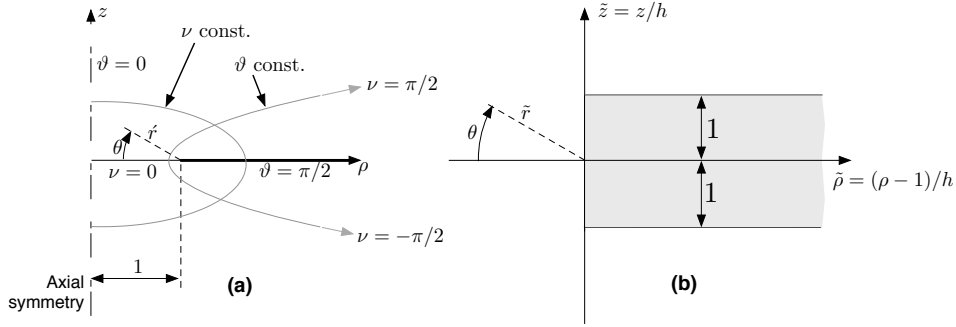}
\caption{Asymptotic regions considered in the small-aspect-ratio limit $h\!\to\!0$. (a) Outer region, where the limiting geometry is that of a circular aperture in a zero-thickness screen. (b) Inner ``corner'' region, where the limiting geometry is two dimensional.}
\label{fig:thin}
\end{center}
\end{figure}

The solution to \eqref{lap thin}--\eqref{far thin} is readily obtained \citep{tuck:1975} by use of separation of variables in oblate spheroidal coordinates $(\nu,\vartheta,\phi)$, which are related to the cylindrical coordinates $(\rho,z,\phi)$ via (see figure \ref{fig:thin}(a))
\refstepcounter{equation}
$$
\rho = \sec \nu \sin \vartheta,\quad z = \tan \nu \cos \vartheta.
\label{coordinates}
\eqno{(\theequation{\mathit{a},\mathit{b}})}
$$
The solution is found to be
\begin{equation}
\label{G h0}
p_0 =  \frac{\nu}{\pi},
\end{equation}
where $-\pi/2\!<\!\nu\!<\!\pi/2$. The far-field behaviour of $p_0$ can be inferred by inverting \eqref{coordinates} at large distances: $\nu\!\sim\! \pm\pi/2\mp1/r + \cdots$ as $r\!\to\!\infty$, respectively for $z\!\gtrless\! 0$. With this relation, \eqref{K in far} readily gives
\begin{equation}
\label{beta h0}
K_0 = 2.
\end{equation}

We shall later require the behaviour of \eqref{G h0} close to the sharp edge of the orifice, which can be obtained by inverting \eqref{coordinates} near $(\nu,\vartheta)\!=\!(0,\pi/2)$. The result is best expressed using polar coordinates ($\acute{r},{\theta}$) in the $\rho\!-\!z$ plane, defined via 
\begin{equation}
\rho-1=-\acute{r}\cos{\theta}, \quad z=\acute{r}\sin{\theta},
\end{equation}
where $-\pi< {\theta}<\pi$. In terms of these coordinates, 
\begin{equation}\label{inner limit of outer}
p_0\sim\frac{\sqrt{2}}{\pi} {\acute r}^{1/2} \sin \frac{{\theta}}{2} + \ldots \quad \text{as} \quad \acute{r}\to0.
\end{equation}
Unsurprisingly, the leading term is proportional to the classical solution of potential flow around a semi-infinite half plane \citep{batchelor:1967}.

\subsection{Non-zero aspect ratio} 
In the limit $h\!\to\!0$, we expand $p$ and $K$ as
\refstepcounter{equation}
$$
p = p_0 + h p_{1} + \ldots, \qquad K = K_{0} + h K_{1} + \ldots,
\label{outer exp thin}
\eqno{(\theequation{\mathit{a},\mathit{b}})}
$$
wherein $p_0$ and $K_0$ are given by \eqref{G h0} and \eqref{beta h0}, respectively. It follows from \eqref{lap}--\eqref{far} that $p_1$ satisfies Laplace's equation
\begin{equation}\label{p1 lap}
\nabla^2 p_1 = 0,
\end{equation}
the inhomogeneous Neumann conditions
\begin{equation}
\label{Neumann condition small h pre}
\pd{p_1}{z} = \mp \frac{\partial^2 p_0}{\partial z^2} \quad \text{at} \quad z=0^{\pm}  \quad (\rho>1)
\end{equation}
and the far-field condition
\begin{equation}\label{p1 far}
p_1\to0 \quad \text{as} \quad r\to\infty. 
\end{equation}

Conditions \eqref{Neumann condition small h pre} were obtained by mapping \eqref{neu} to $z\!=\!0$ by means of a Taylor expansion. Substituting \eqref{G h0}, their explicit form is seen to be
\begin{equation}
\label{Neumann condition small h}
\pd{p_1}{z} = - \frac{1}{\pi}\frac{1}{(\rho^2-1)^{3/2}}\quad \text{at} \quad z=0 \quad (\rho>1),
\end{equation}
implying that $p_1$ is more singular than $p_0$ at the edge $(\rho,z)\!=\!(1,0)$. Now, the problem governing $p_0$ was solved assuming the standard edge condition $|\bnabla p_0|\!=\!O(1/\sqrt{\acute{r}})$ \citep{mittra:1971}. It is clear from \eqref{Neumann condition small h} that $p_1$ cannot also satisfy that condition. Rather, the actual edge condition  shall be determined by matching with an inner ``corner'' region at $O(h)$ distances from the edge. Accordingly, \eqref{outer exp thin}(a) constitutes an outer expansion.

\subsection{Corner region}\label{ssec:corner}
To study the corner region we write 
\begin{equation}
\rho = 1+h\tilde{\rho}, \quad z=h\tilde{z}
\end{equation}
and consider the inner limit $h\!\to\!0$ with the shifted and strained coordinates $(\tilde{\rho},\tilde{z})$ fixed. The pressure in the corner region is written $p(\rho, z)\!=\!\tilde{p}(\tilde{\rho},\tilde{z})$. The limiting geometry of the corner region is depicted in figure \ref{fig:thin}(b); note that it is two dimensional, the fluid domain being that external to a semi-infinite plate.

Expansion \eqref{inner limit of outer}, of the leading-order outer flow close to the sharp edge of the outer geometry, implies that the pressure in the corner region is $O(h^{1/2})$. Thus,
\begin{equation}\label{corner expansion}
\tilde{p} = h^{1/2} \tilde{p}_{1/2} + \ldots \quad \text{as} \quad h\to0. 
\end{equation}
From \eqref{lap}--\eqref{far}, $\tilde{p}_{1/2}$ satisfies the two-dimensional Laplace's equation
\begin{equation}\label{corner lap}
\pd{^2\tilde{p}_{1/2}}{\tilde{\rho}^2} +\pd{^2\tilde{p}_{1/2}}{\tilde{z}^2}=0,
\end{equation}
the Neumann conditions
\refstepcounter{equation}
$$
\pd{\tilde{p}_{1/2}}{\tilde{z}}=0 \quad \text{at} \quad \tilde{z}=\pm 1 \quad (\tilde{\rho}>0) \quad \text{and} \quad 
\pd{\tilde{p}_{1/2}}{\tilde\rho}=0 \quad \text{at} \quad \tilde\rho=0 \quad  (|\tilde{z}|<1)
\label{corner neu}
\eqno{(\theequation{\mathit{a},\mathit{b}})}
$$
and a far-field matching condition. The latter readily follows from \eqref{inner limit of outer} as 
\begin{equation}\label{corner far}
\tilde{p}_{1/2} \sim \frac{\sqrt{2}}{\pi} \tilde{r}^{1/2} \sin \frac{{\theta}}{2} + \ldots, \quad \tilde{r} \to\infty,
\end{equation}
where the polar coordinates ($\tilde{r},{\theta}$), shown in figure \ref{fig:thin}(b), are defined by the relations
\begin{equation}\label{polar corner}
\tilde{\rho}=-\tilde{r}\cos{\theta}, \quad \tilde{z}=\tilde{r}\sin{\theta}
\end{equation}
with $-\pi< {\theta}<\pi$. 

\begin{figure}
   \centering
   \includegraphics[scale=0.55]{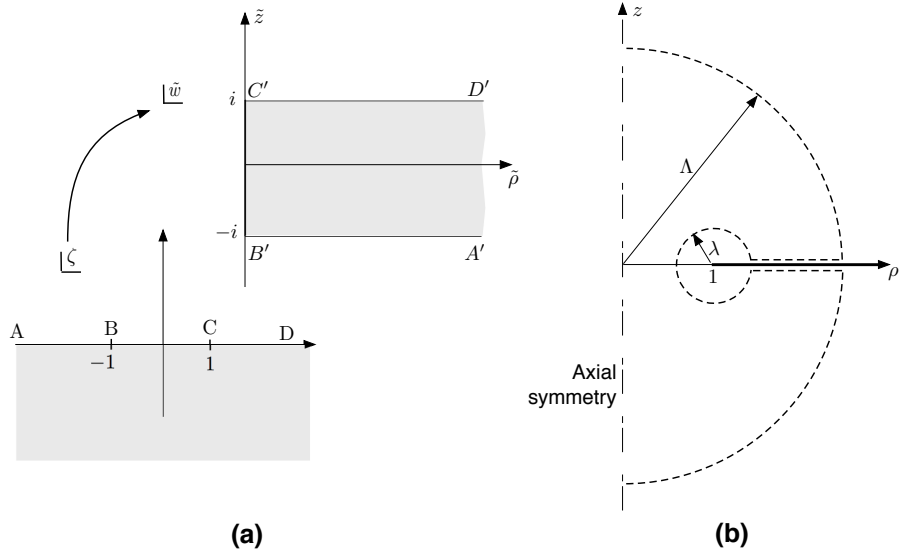}
   \caption{(a) The conformal mapping \eqref{mapping} takes the upper-half $\zeta$ plane to the domain in the $\tilde{w}$ plane exterior to a semi-infinite finite-thickness slab. (b) Integration contour for calculation of $K_1$. }
   \label{fig:mappingkeyhole}
\end{figure}
The problem \eqref{corner lap}--\eqref{corner neu}, together with \eqref{corner far}, is identical to the corner problem solved by \cite{crighton:1973} using conformal mapping. Following these authors, we introduce the complex variable $\tilde{w}\!=\! \tilde{\rho} + i\tilde{z}$ and the conformal mapping 
\begin{equation}\label{mapping}
\tilde{w}  = i +\frac{2}{\pi}\left[\zeta\sqrt{\zeta+1}\sqrt{\zeta-1}-2\log\frac{\sqrt{\zeta+1}+\sqrt{\zeta-1}}{\sqrt{2}}\right] 
\end{equation}
from the upper half of an auxiliary $\zeta$ plane to the corner-region fluid domain. The complex functions in \eqref{mapping} are the principal values with branch cuts along the negative real axis of the $\zeta$ plane. The mapping \eqref{mapping} is readily inverted at large distances:
\begin{equation}\label{mapping inversion}
\zeta \sim \left(\frac{\pi}{2}\right)^{1/2}\tilde{w}^{1/2}+\frac{\log2\pi\tilde{w}+1-i\pi }{{2^{3/2}\pi^{1/2}}\tilde{w}^{1/2}} + \cdots \quad \text{as} \quad \tilde{r}\to\infty,
\end{equation}
wherein the branch cuts of the complex functions are along the positive real axis of the $\tilde{w}$ plane. 

The solution is readily seen to be
\begin{equation}
\label{G 12}
\tilde{p}_{1/2} = \frac{2}{\pi^{3/2}}\operatorname{Re}\zeta,
\end{equation}
namely a uniform flow in the $\zeta$ plane. This solution clearly satisfies Laplace's equation \eqref{corner lap} and the Neumann conditions \eqref{corner neu}; using the leading term in \eqref{mapping inversion}, it is easily checked that the far-field condition \eqref{corner far} is also satisfied. The correction term in \eqref{mapping inversion}, in conjunction with \eqref{polar corner}, can be used to obtain the far-field expansion of \eqref{G 12}:
\begin{equation}\label{asymp corner}
\tilde{p}_{1/2} \sim \frac{\sqrt{2\tilde{r}}}{\pi} \sin\frac{{\theta}}{2}  + \frac{1}{{\pi^2\sqrt{2\tilde{r}}}}\left[\left(1+\ln 2\pi\tilde{r} \right)\sin\frac{{\theta}}{2}-{\theta}\cos\frac{{\theta}}{2}\right] + \cdots \quad \text{as} \quad \tilde{r}\to\infty. 
\end{equation}

\subsection{Leading algebraic correction} 
The appearance of a logarithm in the far-field expansion \eqref{asymp corner} gives rise, via matching, to asymptotic terms of $O(h\ln h)$, in addition to the $O(h)$ terms anticipated in \eqref{outer exp thin}. Following \cite{crighton:1973}, we shall group logarithmically separated terms together, thus ensuring the remainder is of a higher algebraic order in $h$; this is also necessary when applying the Van-Dyke matching rule \cite{van:pert}. With this convention, matching the outer and corner regions provides the requisite edge condition\begin{equation}
\label{Singularity condition}
p_{1} \sim \frac{1}{{\pi^2\sqrt{2\acute{r}}}}\left[\left(1-\ln\frac{h}{2\pi\acute{r}}\right)\sin\frac{\theta}{2}-\theta\cos\frac{\theta}{2}\right]+ \cdots \quad \acute{r} \to0.
\end{equation}

The effective edge condition \eqref{Singularity condition} closes the problem \eqref{p1 lap}--\eqref{p1 far} governing the outer pressure correction $p_1$. In principle, $K_1$ can be extracted from the solution to that problem; in particular, from \eqref{K in far} we have
\begin{equation}\label{p1 K1}
p_1\sim \mp \frac{K_1}{2\pi r}  + o(1/r) \quad \text{as} \quad r\to\infty \quad (z\gtrless0). 
\end{equation}
A reciprocity argument, however, can be used to  obtain $K_1$ without having to solve the problem governing $p_1$ in detail. The argument is based on Green's identity in the form
\begin{equation}
\label{Reciprocal app}
\oint \left(p_{0} \pd{p_{1}}{n} - p_{1}\pd{p_{0}}{n}\right) dA = 0,
\end{equation}
where the integration surface is as shown in figure \ref{fig:mappingkeyhole}(b), with the normal pointing into the domain.  
Consider the integral on the left-hand side of \eqref{Reciprocal app}. Using \eqref{far thin} and \eqref{p1 K1}, we find that the contribution to that integral from the spherical boundary of radius $\Lambda$ approaches $-K_1$ as $\Lambda\to\infty$. The remaining contributions, for $\Lambda=\infty$, are evaluated by allowing the radius $\lambda$ of the cut torus about the sharp edge of the outer geometry to be arbitrarily small. In that limit, the curvature of the torus around the symmetry axis can be neglected. We thereby find
\begin{equation}\label{integrals K1}
K_1 = 2\pi\lim_{\lambda\to0}\left\{{\int_{-\pi}^{\pi}}\left(p_0\pd{p_1}{\acute{r}}-p_1\pd{p_0}{\acute{r}}\right)_{\acute{r}=\lambda}\lambda\, d\theta + 2\int_{1+\lambda}^{\infty}\left(p_0\pd{p_1}{z}\right)_{z=0^+}\rho\,d\rho\right\}.
\end{equation}
Substituting \eqref{G h0}, \eqref{Neumann condition small h} and \eqref{asymp corner}, \eqref{integrals K1} yields
\begin{equation}
K_{1} = -\frac{2}{\pi}\left(\ln\frac{\pi}{h}+1\right).
\end{equation}

\section{Viscous effects}\label{sec:viscous}
\subsection{Problem formulation}\label{ssec:viscousform}
We now extend the dimensionless formulation in \S\ref{sec:form} to include viscous effects. The viscous problem consists of the unsteady (time-harmonic) Stokes equations \refstepcounter{equation}
$$
-i\bu=-\bnabla{p}+\epsilon^2\nabla^2\bu, \quad \bnabla\bcdot\bu=0,\label{unsteady stokes}
\eqno{(\theequation{\mathit{a},\mathit{b}})}
$$
where $\epsilon \!=\!a^{-1}\sqrt{\mu/\varrho\omega}$ represents the thickness of the Stokes boundary layer relative to the orifice radius; the no-slip and impermeability boundary conditions
\begin{equation}\label{no slip}
\bu=\bzero;
\end{equation}
and the far-field condition \eqref{far}. As in the inviscid problem, $p$ is independent of $\phi$ and odd in $z$; the velocity field is written ${\bu}=u\be_{\rho}+w\be_z$, with $u$ odd and $w$  even in $z$.

This section is devoted to deriving approximations to the impedance $\mathcal{Z}$ in the limit $\epsilon\ll1$. In this limit it is convenient to obtain the impedance from $\mathcal{Z}=-i\varrho\omega/aK$, where $K$ is defined, as in the inviscid problem, by the flux integral \eqref{K integral}.

\subsection{Thin Stokes boundary layers}\label{ssec:thinstokes}
We again use matched asymptotics, this time decomposing the fluid domain into an outer region, corresponding to the limit $\epsilon\to0$ with position fixed, and a boundary layer region at $O(\epsilon)$ distances from the orifice walls. In the outer region, we expand the pressure and velocity fields as 
\refstepcounter{equation}
$$
\label{viscous expansions}
{\bu}\sim {\bu}\ub{0} + \epsilon {\bu}\ub{1} + \cdots, \quad {p}\sim {p}\ub{0} + \epsilon {p}\ub{1} + \cdots, \quad \text{as} \quad \epsilon\to0,
\eqno{(\theequation{\mathit{a},\mathit{b}})}
$$
where superscripts will be used to denote orders in $\epsilon$. The normalised Rayleigh conductivity is similarly expanded as 
\begin{equation}
K\sim K\ub{0} + \epsilon K\ub{1} + \cdots, 
\end{equation}
from which the corresponding impedance expansion follows as
\begin{equation}\label{imp K}
\frac{\mathcal{Z}a}{\varrho\omega} \sim -\frac{i}{K\ub{0}}+ \epsilon \frac{iK\ub{1}}{\{K\ub{0}\}^2}+\cdots. 
\end{equation}

From \eqref{unsteady stokes}, both the leading and first-order flow outer fields are inviscid:
\refstepcounter{equation}
$$
\bu\ub{0}=-i\bnabla p\ub{0}, \quad \bnabla\bcdot\bu\ub{0}=0; \qquad \bu\ub{1}=-i\bnabla p\ub{1}, \quad \bnabla\bcdot\bu\ub{1}=0.\label{viscous eq 01}
\eqno{(\theequation{\mathit{a}\!-\!\mathit{d}})}
$$
The far-field conditions \eqref{far} give
\refstepcounter{equation}
$$
\label{viscous far 01}
p\ub{0}\to\pm \frac{1}{2} \quad \text{as} \quad r\to\infty \quad (z\gtrless0); \qquad p\ub{1}\to0 \quad \text{as} \quad r\to\infty.
\eqno{(\theequation{\mathit{a},\mathit{b}})}
$$
Owing to their inviscid nature, the flow fields $\bu\ub{0}$ and $\bu\ub{1}$ cannot satisfy \eqref{no slip}, which stipulates both no-slip and impermeability conditions at the orifice walls. Instead, these flow fields satisfy effective boundary conditions, which are derived in appendix \ref{app:bl} by matching the outer expansions \eqref{viscous expansions} with their counterparts in the boundary layer region; therein, viscous effects enter at leading order and accordingly it is possible to satisfy both no-slip and impermeability.

Unsurprisingly, at leading-order the boundary-layer analysis provides the impermeability boundary condition
\begin{equation}\label{effect bc0}
\bn\bcdot\bnabla p\ub{0}=0.
\end{equation}
It follows that $\bu\ub{0}$ and $K\ub{0}$ correspond to the flow field and normalised Rayleigh conductivity in the inviscid problem, respectively. At the next order, the boundary-layer analysis provides the inhomogeneous Neumann condition
\begin{equation}\label{effect bc}
\bn\bcdot\bnabla p\ub{1}=\frac{1+i}{\sqrt{2}}\nabla_s^2p\ub{0},
\end{equation}
where $\bnabla_s^2$ is the surface Laplacian \citep{van:2007}. This condition represents the displacement of the outer inviscid flow by the Stokes boundary layer. We note that both \eqref{effect bc0} and \eqref{effect bc} apply at the exact geometric boundary of the fluid domain.  

Naively, it may seem that $K\ub{1}$ can be calculated by substituting $\bu\ub{1}$ into  \eqref{K integral}:
\begin{equation}\label{K1 wrong}
K\ub{1}\stackrel{?}{=}-i\iint \bn \bcdot \bu\ub{1}\,  dA.
\end{equation}
Given \eqref{effect bc}, however, the integral on the right hand side is clearly not invariant under the class of permissible surface deformations specified together with \eqref{K integral}. The problem with \eqref{K1 wrong} is that it ignores the $O(1)$ flow in the $O(\epsilon)$-thick Stokes boundary layer, which contributes to the net flux at the same order as $\bu\ub{1}$. This subtlety can be circumvented by first deforming the integration surface in \eqref{K integral} into the intersection of the hemisphere $r\!=\!r^*$ (in $z\!>\!0$) and the fluid domain and then taking $r^*\!\to\!\infty$; it is easy to show that the boundary layer contribution vanishes in that limit. It follows that $K\ub{1}$ can be calculated as 
\begin{equation}\label{viscous K integral}
K\ub{1}{=}\lim_{r^*\to\infty}\iint \bn \bcdot \bnabla p\ub{1}\,  dA,
\end{equation}
where the integration surface is as explained above and we substituted \eqref{viscous eq 01}(c). 

The limit \eqref{viscous K integral} can be evaluated without solving the problem governing $p\ub{1}$ in detail. To this end, we write Green's identity as 
\begin{equation}
\oint\left(p\ub{0}\pd{p\ub{1}}{n}-p\ub{1}\pd{p\ub{0}}{n}\right)\,dA=0,
\end{equation}
where the integral is over the boundary of the fluid volume inside the ball $r\!<\!r^*$. In the limit $r^*\!\to\!\infty$, we use \eqref{viscous far 01}--\eqref{effect bc}, \eqref{viscous K integral} and symmetry to obtain
\begin{equation}\label{K1 before ibp}
{K}\ub{1}=\frac{1+i}{\sqrt{2}}\iint p\ub{0}\nabla_s^2p\ub{0}\,dA,
\end{equation}
where the surface integral is over the entire orifice boundary. The impedance expansion \eqref{imp K} can therefore be written
\begin{equation}\label{imp expansion viscous}
\frac{\mathcal{Z}a}{\varrho\omega} \sim -\frac{i}{K\ub{0}}+ \epsilon \Theta \frac{1- i}{\sqrt{2}}  + \cdots \quad \text{as} \quad \epsilon\to0,
\end{equation}
where the parameter 
\begin{equation}\label{Theta def}
\Theta=\left\{K\ub{0}\right\}^{-2}\iint |\bnabla_s p\ub{0}|^2\,dA
\end{equation}
is a real positive functional of the basic inviscid flow and $\bnabla_s$ is the surface gradient operator \citep{van:2007}; note that we used  integration by parts to arrive at \eqref{Theta def}. 

The parameter $\Theta$ depends solely upon $h$. In what follows, we use the asymptotic solutions obtained in \S\ref{sec:long} and \S\ref{sec:short} to evaluate \eqref{Theta def} in the limits of large and small aspect ratio. To this end, it will be useful to decompose the integral in \eqref{Theta def} as 
\begin{equation}\label{split integral}
\frac{1}{4\pi}\iint |\bnabla_s p\ub{0}|^2\,dA=I_1+I_{2},
\end{equation}
where, using symmetry, 
\refstepcounter{equation}
$$
I_1=\int_0^h\left\{\pd{p\ub{0}}{z}\right\}_{\rho=1}^2\,dz, \quad I_2=\int_1^{\infty}\left\{\pd{p\ub{0}}{\rho}\right\}_{z=h}^2\rho\,d\rho.
\eqno{(\theequation{\mathit{a},\mathit{b}})}
$$

\subsection{Large aspect ratio}
The asymptotic solutions for $p\ub{0}$ obtained in \S\ref{sec:long} and \S\ref{sec:short} are in the form of matched asymptotic expansions. Accordingly, to evaluate the integrals in \eqref{split integral} we shall follow the usual procedure of splitting the range of integration and exploiting the overlap between the matched expansions \citep{hinch:1991}. We begin by considering the limit $h\to\infty$, where the inviscid fluid domain is composed of channel and aperture regions.  

In light of the above, we split $I_{1}$ as
\begin{equation}\label{I1 split}
I_1=\int_0^{h-\lambda}\left\{\pd{p\ub{0}}{z}\right\}_{\rho=1}^2\,dz+\int_{h-\lambda}^{h}\left\{\pd{p\ub{0}}{z}\right\}_{\rho=1}^2\,dz,
\end{equation}
where the arbitrary parameter $\lambda$ satisfies $1\ll \lambda\ll h$. The first integral in \eqref{I1 split} can be evaluated using \eqref{K exp acc} and the channel-region expansion \eqref{Gchannel} 
\begin{equation} \label{i1 int 1}
\int_0^{h-\lambda}\left\{\pd{p\ub{0}}{z}\right\}_{\rho=1}^2\,dz\sim \frac{h-\lambda}{4(h+l)^2}+e.s.t. 
\end{equation}
The second integral is evaluated using \eqref{K exp acc} and the aperture-region expansion \eqref{long channel exp},
\begin{equation}
\int_{h-\lambda}^{h}\left\{\pd{p\ub{0}}{z}\right\}_{\rho=1}^2\,dz\sim \frac{1}{4(h+l)^2}\int_{-\lambda}^0\left\{\pd{g}{\bar{z}}\right\}_{\rho=1}^2\,d\bar{z} +e.s.t.
\end{equation}
While the latter integral does not converge as $\lambda\to\infty$, using \eqref{g channel} it can be written 
\begin{equation} \label{i1 int 2}
\int_{-\lambda}^0\left\{\pd{g}{\bar{z}}\right\}_{\rho=1}^2\,d\bar{z}=\int_{-\lambda}^0\left(\left\{\pd{g}{\bar{z}}\right\}_{\rho=1}^2-1\right)\,d\bar{z}+\lambda.
\end{equation}
The integrand of the integral on the right-hand  side of \eqref{i1 int 2} decays exponentially as $\bar{z}\!\to\!-\infty$; replacing the lower integration limit by $-\infty$ therefore introduces only an exponentially small error. When \eqref{i1 int 1}--\eqref{i1 int 2} are summed according to \eqref{I1 split}, the dependence upon the arbitrary parameter $\lambda$ is eliminated. We thereby find
\refstepcounter{equation}
$$
\label{I1 long as}
I_1\sim \frac{h+b_1}{4(h+l)^2}+e.s.t. \quad \text{as} \quad h\to\infty, \quad  b_1=\int_{-\infty}^0\left(\left\{\pd{g}{\bar{z}}\right\}_{\rho=1}^2-1\right)\,d\bar{z},
\eqno{(\theequation{\mathit{a},\mathit{b}})}
$$
where $b_1$ is a constant, a quadrature of the canonical aperture problem defined in \S\S\ref{ssec:aperture}.

The integral $I_2$ in \eqref{split integral} is readily evaluated as it only involves the aperture region. Using \eqref{Gchannel} and \eqref{K exp acc}, 
\refstepcounter{equation}
$$\label{I2 long as}
I_2\sim \frac{b_2}{4(h+l)^2}+e.s.t., \quad b_2=\int_1^{\infty}\left\{\pd{g}{\rho}\right\}_{\bar{z}=0}^2\rho\,d\rho,
\eqno{(\theequation{\mathit{a},\mathit{b}})}
$$
where $b_2$ is a constant, another quadrature of the canonical aperture problem. 

Substituting \eqref{K exp acc}, \eqref{I1 long as} and \eqref{I2 long as} into \eqref{Theta def} and \eqref{split integral}, we find
\begin{equation}
\Theta\sim \frac{4}{\pi}(h+l_{v}) + e.s.t. \quad \text{as} \quad h\to\infty,
\end{equation}
where $l_{v}=b_1+b_2$. The constants $b_{1}$ and $b_{2}$ [cf.~\eqref{I1 long as}(b) and \eqref{I2 long as}(b)] are evaluated using the same numerical solution of the canonical aperture problem used in \S\ref{sec:long} to calculate $l$. We thereby find $l_{v}\approx 0.91$. 

\subsection{Small aspect ratio}
Consider now the small-aspect-ratio limit $h\!\to\!0$, where the inviscid fluid domain is composed of an outer region and an inner corner region. 

In this limit, the integral $I_{1}$ solely involves the corner region. Thus,  substituting the corner expansion \eqref{corner expansion},
\begin{equation}\label{I1 short corner}
I_1\sim \int_0^1\left(\pd{\tilde{p}_{1/2}}{\tilde{z}}\right)^2_{\tilde{\rho}=0}\,d\tilde{z}  \quad \text{as} \quad h\to0,
\end{equation} 
where here and for the remainder of this subsection the relative asymptotic error is of an unspecified algebraic order. The integral in \eqref{I1 short corner} can be evaluated using the closed-form solution \eqref{G 12} for $\tilde{p}_{1/2}$, which relies on the conformal mapping \eqref{mapping}. Using a change of variables from $\tilde{z}$ to $\tau\!=\!\operatorname{Re}\zeta$, we find
\begin{equation}\label{I1 short as}
I_1\sim \int_0^1\left(\pd{\tilde{p}_{1/2}}{\tilde{z}}\right)^2_{\tilde{\rho}=0}\,d\tilde{z}=  \int_0^1\left\{\frac{d}{d\tau} (\tilde{p}_{1/2})_{\tilde{\rho}=0}\right\}^2\left(\frac{d\tilde{z}}{d\tau}\right)^{-1}\,d\tau=\frac{1}{2\pi}.
\end{equation}

Both the corner and outer regions contribute to $I_2$ at leading order. We therefore split that integral as 
\begin{equation}\label{I2 small}
I_2 = \int_{1}^{1+\lambda}\left(\pd{p}{\rho}\right)^2_{z=h}\rho\,d\rho+\int_{1+\lambda}^{\infty}\left(\pd{p}{\rho}\right)^2_{z=h}\rho\,d\rho,
\end{equation}
where the arbitrary parameter $\lambda$ is now assumed to satisfy $h\!\ll\! \lambda\!\ll\!1$. Substituting the corner expansion \eqref{corner expansion} into the first integral in \eqref{I2 small} gives
\begin{equation} 
\int_{1}^{1+\lambda}\left(\pd{p}{\rho}\right)^2_{z=h}\rho\,d\rho\sim \int_0^{\lambda/h}\left(\pd{\tilde{p}_{1/2}}{\tilde{\rho}}\right)^2_{\tilde{z}=1}\,d\tilde{\rho},
\end{equation}
To evaluate the latter integral we again use the conformal mapping solution \eqref{G 12} together with a change variables from 
$\tilde{\rho}$ to $\tau=\operatorname{Re}\zeta$: 
\begin{equation}\label{I2 small cont1}
\int_0^{\lambda/h}\left(\pd{\tilde{p}_{1/2}}{\tilde{\rho}}\right)^2_{\tilde{z}=1}\,d\tilde{\rho}=\int_1^t\left\{\frac{d}{du}\left(\tilde{p}_{1/2}\right)_{\tilde{z}=1}\right\}^2\left(\frac{d\tilde{\rho}}{du}\right)^{-1}\,du\sim \frac{1}{2\pi^2}\ln \frac{2\pi\lambda}{h},
\end{equation}
wherein, from \eqref{mapping}, $t\sim (\pi\lambda/2h)^{1/2}$ as $\lambda/h\to\infty$. 
The second integral in \eqref{I2 small} is readily evaluated by substituting the outer solution \eqref{G h0}:
\begin{equation}\label{I2 small cont2}
\int_{1+\lambda}^{\infty}\left(\pd{p}{\rho}\right)^2_{z=h}\rho\,d\rho\sim \int_{1+\lambda}^{\infty}\left(\pd{p_0}{\rho}\right)^2_{z=0^+}\rho\,d\rho = \frac{1}{2\pi^2}\ln \frac{1}{2\lambda}.
\end{equation}
Summing \eqref{I2 small cont1} and \eqref{I2 small cont2} according to \eqref{I2 small}, the dependence upon $\lambda$ disappears:
\begin{equation}\label{I2 short as}
I_{2} \sim \frac{1}{2\pi^2}\ln \frac{\pi}{h}   \quad \text{as} \quad h\to0.
\end{equation} 

Substituting \eqref{beta h0}, \eqref{I1 short as} and \eqref{I2 short as} into \eqref{Theta def} and \eqref{split integral} gives
\begin{equation}\label{theta short again}
\Theta\sim \frac{1}{2}\left(\frac{1}{\pi}\ln\frac{\pi}{h}+1\right) + \cdots \quad \text{as} \quad h\to0.
\end{equation}

\subsection{The case $h=O(\epsilon)$}
According to \eqref{imp expansion viscous} and \eqref{theta short again}, the acoustic resistance $\operatorname{Re}\mathcal{Z}$ diverges as $h\to0$.
When $h$ is so small as to be commensurate with $\epsilon$, however, the inner corner region analysed in \S\ref{sec:short} can no longer be assumed inviscid, even to leading order in $\epsilon$; accordingly, it is no longer permissible  to consider the limits $\epsilon\to0$ and $h\!\to\!0$ in succession. 

To analyse this special case, we revisit the limit $\epsilon\to0$, this time with $h/\epsilon$, rather than $h$, fixed. The outer flow coincides at $O(1)$ with the basic inviscid flow in the case $h\!=\!0$ and hence $K\!\sim \!2$. The outer flow at $O(\epsilon)$ remains inviscid, with the pressure field satisfying the effective boundary condition
\begin{equation}\label{neu extreme}
\pd{p\ub{1}}{z} =\pm\left( \frac{1+i}{\sqrt{2}} + \frac{h}{\epsilon}\right)\nabla_s^2p\ub{0} \quad \text{at} \quad z=0^{\pm} \quad (\rho>1); 
\end{equation}
note that this condition, which can be most easily inferred by mapping \eqref{effect bc} to $z\!=\!0$ by means of a Taylor expansion, combines the displacement effects of the plate thickness and Stokes boundary layer. In order to close the problem governing $p\ub{1}$ it is necessary to match the outer region with an inner corner region at distances $O(h)$, or $O(\epsilon)$, from the orifice edge. The leading-order flow field in that region satisfies a two-dimensional unsteady Stokes problem, with no-slip and impermeability boundary conditions prescribed on the corner boundary and a far-field matching condition similar to \eqref{corner far}.

Solving the unsteady-Stokes corner problem is an essential step towards deriving an algebraically accurate approximation for $(K\!-\!2)$, on a par with those previously obtained in both the inviscid and viscous small-aspect-ratio limits. It is not clear, however, how to make analytical progress in that direction; in the special case $h\!=\!0$, the Wiener--Hopf method used by Alblas \cite{alblas:1957} to analyse viscous effects on plane-wave diffraction from a sharp edge may be useful. Unfortunately, a numerical solution of the corner problem would inconveniently depend on the ratio $h/\epsilon$.  

To confirm that the resistance does not actually diverge as $h\!\to\!0$, it suffices to derive a crude --- logarithmically accurate --- approximation, which can be readily deduced using \eqref{neu extreme} and the assumed asymptotic overlap between the boundary layer and corner regions. Thus, using a reciprocal argument and integration surface as in \S\ref{sec:short}, we obtain
\begin{equation}
K\sim 2 -\frac{2\epsilon}{\pi}\left(\frac{1+i}{\sqrt{2}}+\frac{h}{\epsilon}\right)\ln\frac{1}{\epsilon}+O(\epsilon),
\end{equation}
which is \eqref{intro special} when rewritten as impedance.

\section*{Acknowledgements.} 
O.S. acknowledges funding from the Engineering and Physical Sciences Research Council UK (award no.~EP$/$R041458$/$1).

\appendix
\section{Electrical analogy of the inviscid problem}\label{app:anal}
Rayleigh's acoustic conductivity $\mathcal{K}$ can be thought of as the inverse of an impedance defined as ${\Phi}/\mathcal{Q}$, instead of $\mathcal{P}/\mathcal{Q}$, where ${\Phi}$ is the velocity potential difference across the perforation. (Clearly, this interpretation only makes sense in the inviscid regime.) As noted in the introduction, 
Rayleigh's use of the term conductivity in this context appears to be motivated by a mathematical analogy between the inviscid problem and that of calculating the effective electrical conductivity $\mathcal{K}_E=\mathcal{I}/\mathcal{V}$ of a perforation in a perfectly insulating plate, $\mathcal{V}$ and $\mathcal{I}$ respectively denoting the voltage across and current through the perforation \citep{rayleigh:1871,rayleigh:1896}.

It is beneficial to make this analogy explicit. Ohm's law in the medium external to the plate is 
\begin{equation}\label{ohm}
\bi=-\sigma \bnabla^*\varphi,
\end{equation}
where $\varphi$ denotes the electric potential, $\sigma$ is the conductivity of the medium and $\bnabla^*$ is the dimensional gradient (in contrast to the normalised gradient $a\bnabla^*=\bnabla$ used throughout the paper). The problem governing $\varphi$ consists of the constitutive relation \eqref{ohm}, the charge-conservation equation
\begin{equation}\label{charge}
\bnabla^*\bcdot\bi=0,
\end{equation}
the insulating boundary condition
\begin{equation}\label{insulating}
\bn\bcdot\bi=0
\end{equation}
and the far-field conditions
\begin{equation}\label{voltage}
\varphi \to \pm \frac{1}{2}\mathcal{V}  \quad \text{as} \quad r\to\infty \quad (z\gtrless0).
\end{equation} 
Clearly, the ratio $\varphi/\mathcal{V}$ satisfies the same problem as $p$ in the inviscid case (cf.~\S\ref{sec:form}). It readily follows that 
\begin{equation}\label{analogy}
\mathcal{K}_E=\sigma \mathcal{K}. 
\end{equation}

Thus, the asymptotic approximations \eqref{long inv} and \eqref{short inv} derived in the context of inviscid acoustics are readily translated to the electrical problem. Similar analogies with problems of heat and mass diffusion, as well as other potential problems (collectively known as ``blockage problems'') can be easily established \citep{tuck:1975,sherwood:2014}. Moreover, the same approach used herein to incorporate viscous effects may be relevant when considering physical generalisations of these analogous problems.

\section{Stokes boundary layers}\label{app:bl}
\subsection{Surface parameterisation and boundary-fitted coordinates}
We here derive the effective boundary condition \eqref{effect bc} by matching the outer expansions \eqref{viscous expansions} with a Stokes boundary layer expansion valid at $O(\epsilon)$ distances from the orifice walls. In the present paper the orifice is composed of relatively simple flat and cylindrical rigid surfaces; for generality, we shall analyse the boundary layer assuming an arbitrary smooth surface. Let this surface be locally parameterised as $\bx=\by(\xi,\eta)$, where $(\xi,\eta)$ are orthogonal surface coordinates with unity metric coefficients, \textit{viz.},
\refstepcounter{equation}
$$
\pd{\by}{\xi}\bcdot\pd{\by}{\eta}=0, \quad \left|\pd{\by}{\xi}\right|=\left|\pd{\by}{\eta}\right|=1.
\eqno{(\theequation{\mathit{a},\mathit{b}})}
$$
The fluid domain in a neighbourhood of the surface is conveniently described in terms of curvilinear ``boundary-fitted'' coordinates $(\xi,\eta,\chi)$, with corresponding unit vectors $(\be_{\xi},\be_{\eta},\be_{\chi})$; these coordinates extend the surface coordinates $(\xi,\eta)$ into the fluid domain such that $\bx=\by(\xi,\eta)+\bn\chi$, $\bn$ being the outward unit normal at $(\xi,\eta)$ and $\chi>0$. 

In this appendix, $p$ and $\bu$ are considered to be  functions of $(\xi,\eta,\chi)$:
\refstepcounter{equation}
$$
p=p(\xi,\eta,\chi), \quad \bu=\be_{\xi}u(\xi,\eta,\chi)+\be_{\eta}v(\xi,\eta,\chi)+\be_{\chi}w(\xi,\eta,\chi).
\eqno{(\theequation{\mathit{a},\mathit{b}})}
$$
In accordance with \S\ref{sec:viscous}, we assume that $p$ and $\bu$ satisfy the unsteady Stokes equations \eqref{unsteady stokes} and the no-slip and impermeability boundary conditions \eqref{no slip}. The former generally possess a complicated form when written in terms of boundary-fitted coordinates; in contrast, the latter simply read
\begin{equation}\label{fitted bc}
u=v=w=0 \quad \text{at} \quad \chi=0.
\end{equation}
For later reference, we note that $\be_{\chi}=\bn$ at the boundary $\chi=0$ and hence  
\refstepcounter{equation}
$$
\bn\bcdot\bu = w, \quad \bn\bcdot\bnabla p =\pd{p}{\chi} \quad \text{at} \quad \chi=0. 
\label{simple}
\eqno{(\theequation{\mathit{a},\mathit{b}})}
$$

\subsection{Outer expansions}
It is helpful to restate the outer expansions \eqref{viscous expansions} as 
\begin{equation}\label{bl outer expansions}
{g}(\xi,\eta,{\chi})\sim {g}\ub{0}(\xi,\eta,{\chi})+\epsilon{g}\ub{1}(\xi,\eta,{\chi})+ \cdots \quad \text{as} \quad \epsilon\to0,
\end{equation}
where here $g$ stands for either $p,u,v$ or $w$. The outer fields satisfy the inviscid equations \eqref{viscous eq 01} at $O(1)$ and $O(\epsilon)$, respectively. We shall only need the transverse momentum balances on the boundary
\refstepcounter{equation}
$$
\label{transmom bl}
\pd{p\ub{0}}{\chi}=iw\ub{0}, \quad \pd{p\ub{1}}{\chi}=iw\ub{1}\quad \text{at} \quad \chi=0, 
\eqno{(\theequation{\mathit{a},\mathit{b}})}
$$
the form of which are easily inferred using \eqref{simple}.
 
\subsection{Boundary layer analysis and matching}
The Stokes boundary layer adjacent to the surface corresponds to the inner limit $\epsilon\to0$ with the strained normal coordinate $\tilde{\chi}=\chi/\epsilon$ fixed. We write the pressure and velocity fields in this region as 
\refstepcounter{equation}
$$
p=\tilde{p}(\xi,\eta,\tilde{\chi}), \quad \bu=\be_{\xi}\tilde{u}(\xi,\eta,\tilde{\chi})+\be_{\eta}\tilde{v}(\xi,\eta,\tilde{\chi})+\be_{\chi}\tilde{w}(\xi,\eta,\tilde{\chi}),
\eqno{(\theequation{\mathit{a},\mathit{b}})}
$$
hence the no-slip and impermeability boundary conditions \eqref{fitted bc} become:
\begin{equation}\label{bl bc}
\tilde{u}=\tilde{v}=\tilde{w}=0 \quad \text{at} \quad \tilde{\chi}=0.
\end{equation}
To proceed, we expand the fields $\tilde{p},\tilde{u},\tilde{v}$ and $\tilde{w}$ according to the generic inner expansion
\begin{equation}
\tilde{g}(\xi,\eta,\tilde{\chi})\sim \tilde{g}\ub{0}(\xi,\eta,\tilde{\chi})+\epsilon\tilde{g}\ub{1}(\xi,\eta,\tilde{\chi})+ \cdots \quad \text{as} \quad \epsilon\to0.
\end{equation}

In the boundary layer, the form of the unsteady Stokes equations \eqref{unsteady stokes} can be simplified by expanding the scale factors of the strained boundary-fitted coordinates $(\xi,\eta,\tilde{\chi})$ as $\epsilon\to0$. In particular, the leading-order balances of the momentum equations are 
\refstepcounter{equation}
$$
\pd{\tilde{p}\ub{0}}{\tilde{\chi}}=0, \quad 
\pd{^2\tilde{u}\ub{0}}{\tilde{\chi}^2}+i\tilde{u}\ub{0}=\pd{\tilde{p}\ub{0}}{\xi}, \quad \pd{^2\tilde{v}\ub{0}}{\tilde{\chi}^2}+i\tilde{v}\ub{0}=\pd{\tilde{p}\ub{0}}{\eta}.\label{bl mom}
\eqno{(\theequation{\mathit{a}\!-\!\mathit{c}})}
$$
It follows from \eqref{bl mom}(a) that $\tilde{p}\ub{0}$ is independent of $\chi$; asymptotic matching thus gives
\begin{equation}\label{bl matching p0}
\tilde{p}\ub{0}=p\ub{0}(\xi,\eta,0). 
\end{equation}
Next, integrating \eqref{bl mom}(b) and \eqref{bl mom}(c) with respect to $\tilde{\chi}$ and using \eqref{bl bc} at $O(1)$ gives 
\begin{equation}\label{uv solution}
\left\{\tilde{u}\ub{0},\tilde{v}\ub{0}\right\}=-i\left\{\pd{\tilde{p}\ub{0}}{\xi},\pd{\tilde{p}\ub{0}}{\eta}\right\}\left(1-e^{-\frac{1-i}{\sqrt{2}}\tilde{\chi}}\right).
\end{equation}
Consider next the continuity equation \eqref{unsteady stokes}(b), which can be written
\begin{equation}\label{bl cont}
\pd{}{\xi}\left\{\left[1+O(\epsilon)\right]\tilde{u}\right\}+
\pd{}{\eta}\left\{\left[1+O(\epsilon)\right]\tilde{v}\right\}+\frac{1}{\epsilon}\pd{}{\tilde{\chi}}\left\{\left[1+O(\epsilon)\right]\tilde{w}\right\}=0.
\end{equation}
From the $O(1/\epsilon)$ balance of \eqref{bl cont} we find, using \eqref{bl bc},
\begin{equation}\label{w is zero}
\tilde{w}\ub{0}\equiv 0.
\end{equation}
Thus, leading-order matching of the inner-outer expansions of $w$ gives
\begin{equation}
w\ub{0}= 0 \quad \text{at} \quad \chi=0,
\end{equation}
which, with \eqref{simple} and \eqref{transmom bl}(a), provides the leading-order effective boundary condition
\begin{equation}
\bn\bcdot\bnabla p\ub{0} = 0\quad \text{at} \quad \chi=0. 
\end{equation}

With \eqref{w is zero}, the $O(1)$ balance of \eqref{bl cont} becomes
\begin{equation}
\pd{\tilde{w}\ub{1}}{\tilde{\chi}}=-\left(\pd{\tilde{u}\ub{0}}{\xi}+\pd{\tilde{v}\ub{0}}{\eta}\right).
\end{equation}
Substitution of \eqref{uv solution}, followed by integration with respect to $\tilde{\chi}$ and using the $O(\epsilon)$ balance of \eqref{bl bc}, 
\begin{equation}\label{w1}
\tilde{w}\ub{1}=i\left\{\pd{^2\tilde{p}\ub{0}}{\xi^2}+\pd{^2\tilde{p}\ub{0}}{\eta^2}\right\}\left[\tilde{\chi}+\frac{1+i}{\sqrt{2}}\left(e^{-\frac{1-i}{\sqrt{2}}\tilde{\chi}}-1\right)\right].
\end{equation}
Noting the matching condition \eqref{bl matching p0}, the term in the curly brackets can be interpreted as the surface Laplacian \citep{van:2007}
\begin{equation}
\nabla_s^2p\ub{0}=\pd{^2{p}\ub{0}}{\xi^2}+\pd{^2{p}\ub{0}}{\eta^2} \quad \text{at} \quad \chi=0.
\end{equation}

To obtain the requisite effective boundary condition satisfied by the $O(\epsilon)$ outer pressure, we match the inner-outer expansions of $w$ to higher order. To this end, we write the two-term outer expansion of $w$ in terms of the inner coordinate $\tilde{\chi}$ and expand as $\epsilon\to0$:
\begin{equation}
w=\left\{w\ub{0}\right\}_{\chi=0}+\epsilon\left[\tilde{\chi}\left\{\pd{w\ub{0}}{\chi}\right\}_{\chi=0}+\left\{w\ub{1}\right\}_{\chi=0}\right] + \cdots
\end{equation}
 Since $w\ub{1}=\bu\ub{1}\bcdot\bn$ at $\xi=0$, matching with \eqref{w is zero} and \eqref{w1} gives
  \begin{equation}
w\ub{1}=-i\nabla_s^2p\ub{0}\frac{1+i}{\sqrt{2}} \quad \text{at} \quad \chi=0,
 \end{equation}
which, using \eqref{simple} and \eqref{transmom bl}(b), gives 
 \begin{equation}
\bn\bcdot\bnabla p\ub{1} =\nabla_s^2p\ub{0}\frac{1+i}{\sqrt{2}} \quad \text{at} \quad \chi=0,
 \end{equation}
in agreement with \eqref{effect bc}.

\bibliography{refs}

\end{document}